\documentstyle[multicol,aps,epsf,psfig,graphics]{revtex}

\def\CF{{\cal{F}}}

\def\CN{{\cal{N}}}

\def\x{\vec{x}}
\def\e{\hat{e}}
\def\partd#1#2{{\frac{\partial #1}{\partial #2}}}

\begin{document}
\draft
\title{A thermodynamically reversible generalization of {Diffusion
Limited Aggregation}}
\author{Raissa M. D'Souza}
\address{Department of Physics, Massachusetts Institute of Technology,
Cambridge, Massachusetts 02139}
\author{Norman H. Margolus}
\address{Center for Computational Science, Boston University, Boston,
Massachusetts 02215\\
Artificial Intelligence Laboratory, Massachusetts Institute of
Technology, Cambridge, Massachusetts 02139}
\date{\today}
\maketitle
\begin{abstract}
        We introduce a lattice gas model of cluster growth via the
diffusive aggregation of particles in a closed system obeying a local,
deterministic, microscopically reversible dynamics. This model
roughly corresponds to placing the irreversible Diffusion Limited
Aggregation model (DLA) in contact with a heat bath.  Particles release
latent heat when aggregating, while singly connected cluster members
can absorb heat and evaporate. The heat bath is initially empty, hence
we observe the flow of entropy from the aggregating gas of particles
into the heat bath, which is being populated by diffusing heat tokens.
Before the population of the heat bath stabilizes, the
cluster morphology (quantified by the fractal dimension) is similar to
a standard DLA cluster.  The cluster then gradually anneals, becoming
more tenuous, until reaching configurational equilibrium when the
cluster morphology resembles a quenched
branched random polymer.  As the microscopic dynamics is invertible,
we can reverse the evolution, observe the inverse flow of heat and
entropy, and recover the initial condition.  This simple system
provides an explicit example of how macroscopic dissipation and
self-organization can result from an underlying microscopically
reversible dynamics.     

        We present a detailed description of the dynamics for the
model, discuss the macroscopic limit, and give predictions for the
equilibrium particle densities obtained in the mean field limit.
Empirical results for the growth are then presented, including the
observed equilibrium particle densities, the temperature of the system,
the fractal dimension of the growth clusters, scaling
behavior, finite size effects, and the approach to equilibrium.  We
pay particular attention to the temporal behavior of the growth
process and show that the relaxation to the maximum entropy state is
initially a rapid non-equilibrium process, then subsequently it is a
quasistatic process with a well defined temperature. 

\end{abstract}

\begin{multicols}{2}\setlength{\columnwidth}{\hsize}

\section{Microscopic reversibility and pattern formation} \label{intro}

Pattern formation is an intrinsically
dissipative process\cite{Prig}, however the laws of
physics are microscopically reversible: there is no
dissipation at the microscopic scale.  In this paper we
describe a simple system which organizes into patterns through 
microscopically reversible dynamics, hence it also models how 
dissipation arises (i.e., how information flows between the
macroscopic and the microscopic degrees of freedom). This system
provides a clear example of how to reconcile the macroscopic
irreversibility that gives rise to patterns with the microscopic
reversibility adhered to by physical processes.  Motivated by the
desire to understand this general issue,
we study specific details of the model, focusing on 
transitions in the resulting growth morphology and the approach to
thermodynamic equilibrium.

We have previously observed several examples of reversible
cellular automata dynamics which produce large scale order through
microscopically reversible
dynamics\cite{cambook,Margolus-feynlec,dsouza-same3d}.   
In contrast other research in the field of pattern formation has 
focused on irreversible microscopic mechanisms, with examples ranging
from crystal growth\cite{Vold,KKRSOS}, to Turing patterns
in chemical reactions\cite{Pearson_turing}, to patterns formed by 
growing bacterial colonies\cite{Ben-Jac_bacsnowflakes}, to kinetic
growth problems\cite{Eden-model,WittSand-dla}.   

Here we model cluster growth by reversible aggregation (RA),
generalizing the irreversible Diffusion Limited Aggregation model
(DLA)\cite{WittSand-dla} to include contact with a heat bath.
Particles, which are initially diffusing on a two-dimensional lattice,
stick upon first contacting a cluster member and release heat which
then diffuses about a superimposed lattice representing the heat bath.
The two subsystems exchange only heat and
together form a closed system. The release of heat
transfers entropy from the aggregating system (which is becoming
ordered) into the heat bath (which was initially empty).  
When the heat bath is nearly empty the model is essentially equivalent
to the canonical DLA formulation (analogous to a supercooled gas
crystalizing in a far from equilibrium situation).  Hence the RA
growth cluster initially resembles a typical DLA cluster. As the heat
bath becomes populated, singly connected cluster members are able to
absorb heat and evaporate. As the effect of evaporation becomes
significant the RA and DLA models diverge.  The RA dynamics is exactly
invertible: at any point we can invert the dynamics and run backwards,
observing the flow of heat from the heat bath back into the
gas-crystal system until we recover the exact initial condition.

The population levels of the heat bath and of the aggregate initially
grow linearly in time, quickly reaching stable values
which remain very nearly constant for the remainder of the evolution.
The energy of each subsystem is a function only of the population
levels, independent of the physical configuration of the
particles. Hence, once the population levels stabilize, the rate of
energy exchange (which is entirely in the form of heat) between the
heat bath and the gas-aggregate system becomes so slow that we can
characterize the subsequent dynamics as a quasistatic process, with a
well defined temperature at all times.   

The aggregate mostly forms while the heat bath is at a lower
temperature than in the quasistatic steady-state. Hence, after the
population levels stabilize, the cluster 
slowly anneals. The cluster morphology, which initially
resembles a DLA cluster, gradually becomes more spread out and tortuous,
until it ultimately resembles a branched polymer with 
quenched randomness.
The two timescales that characterize the growth process are
separated by two orders of magnitude.  Initially, the population levels
quickly reach a quasistatic steady-state.  Subsequently,
the aggregate slowly anneals until reaching the ensemble of
configurations corresponding to the highest entropy macrostate 
(the branched polymer). 

Aside from insight into microscopically reversible mechanisms that
give rise to macroscopic patterns, the 
development of invertible dynamics and algorithms has technological
significance in pushing down the barrier to atomic scale
computing.  Each bit of information erased at temperature $T$
releases at least $T \Delta S = k_B T \ln 2$ units of heat into the
environment\cite{Landauer-diss61}. Heat is created in proportion to
the volume of the computer, yet heat leaves the computer only in
proportion to the surface area. Hence, as logic gate density
in computers increases, the use of an invertible dynamics (which
does not erase information and hence does not need to produce heat)
will be required to keep the mechanical parts from burning 
up\cite{Benn-thermo,mpf-foresight}.
{From} a more pedagogical viewpoint, discrete computer models of
reversible microscopic dynamics provide a laboratory for studying 
non-equilibrium statistical mechanics and the approach to
equilibrium.  These models let us explore physically plausible dynamics  
for non-equilibrium systems (i.e., discrete dynamics which are
microscopically reversible and thus automatically obey Liouville's
Theorem).  A particularly instructive example of 
this approach is the formulation of a dynamical Ising 
model\cite{Creutz-annphys86,Pomeau-jphysa84,Vichniac-physd84}.
However, more widely used in physics are discrete, reversible models
of fluid flow such as the HPP and FHP lattice
gases\cite{HPP-pra76,FHP-prl86}. For a recent discussion of modeling
physical phenomena with reversible 
computer models see Ref. \cite{Margolus-feynlec}. For a recent
discussion of macroscopic irreversibility and microscopic
reversibility see Ref. \cite{Lebowitz}.  For a recent discussion of
techniques for the explicit construction of reversible models in
statistical mechanics see Ref.~\cite{LevesVerl-jstatphys93}; but note
that closely related techniques were discussed in the early 1980's by
Fredkin (as discussed in Ref.~\cite{nhm-physlike}).

The initial sections of this manuscript describe 
our model; the middle contain a mathematical formulation of the model;
the final, the empirical results. Specifically, Sec. \ref{sec.mod-agg}
describes the detailed dynamics, including the subtleties of
constructing an invertible dynamical model and implementation
issues. In Sec. \ref{analytics} we discuss the macroscopic limit of
an analytic formulation of the model and establish the
reaction-diffusion equations describing the system.  In
Sec. \ref{meanfield} we treat the reaction-diffusion equations in the
mean field limit and compare predictions for equilibrium densities
of particles to empirical measurements.  Empirical measurements of
temperature are presented in Sec. \ref{temp}, with emphasis on the
quasistatic nature of the annealing portion of the growth process.
In Sec. \ref{frac-dim-results} we study the evolution of the fractal
dimension of the clusters and thus quantify
the change in growth morphology as the clusters relax to the maximum 
entropy state.  We conclude with a discussion of limitations and
possible modifications of our model. 

\section{Modeling aggregation} \label{sec.mod-agg}

\subsection{Diffusion Limited Aggregation}
Diffusion Limited Aggregation (DLA) \cite{WittSand-dla} is a
conceptually simple model which serves as a paradigm for some aspects
of kinetic growth phenomena. Several comprehensive reviews of DLA
have been written. In particular see Ref.~\cite{Stanley-dla-rev} for a
clear presentation of the basics, Ref.~\cite{Langer-rmp80} for details
on physical mechanisms,  and Ref.~\cite{ErzanPietVesp-rmp95} for a
review of real-space renormalization group approaches to DLA.

The typical scenario for DLA begins with a vacant two-dimensional
lattice initialized with a single stationary seed particle, which is the
nucleation site for a growth cluster.  Moving particles are introduced
from the edges of the lattice, following random walks along the
lattice sites. When a moving particle lands on a site adjacent to  
a stationary seed particle (an active site) it sticks (i.e.,
the moving particle aggregates and becomes a stationary seed
particle).  The frozen aggregate particles 
constitute the solid (crystal) phase, and the moving particles
constitute the gas phase. Aggregation hence consists of a particle
undergoing an irreversible transition from gas to solid. Gas
particles are usually introduced in a serial manner: only one gas
particle is diffusing at a time.  However, to take advantage of
parallel computational resources, parallel models of DLA
have been studied in which multiple particles are diffusing at
once\cite{Voss-jstatp84,Naga-jphysjpn92}. In the dilute particle
limit these models recover the serial DLA model exactly. 

With the first aggregation event the DLA cluster grows from one single 
to two adjacent sites. The presence of the second cluster member   
eliminates certain paths along which random walkers could approach the
first, 
with the effect that the probability of sticking at either end (tip)
of the cluster is enhanced, whereas the probability of sticking 
along the edge of the cluster is reduced. As particles continue to
aggregate creating new cluster tips and edges, the probability to
stick at the tips continually outweighs the probability to stick along
the edges. This leads to branching.  A second influence on the
growth morphology comes from shadowing: the probability for a particle to
diffuse into the center of the growth cluster before encountering an
active site becomes negligible as the cluster grows in size. Hence the
outer tips grow most rapidly. As a result the growth aggregate rapidly
assumes a bushy and branching, random fractal structure, resembling
frost on a window pane, the branching of neurons, and many
other branched structures found in nature. 

\subsection{Reversible Aggregation}
Our goal is to introduce a reversible, deterministic model of growth
by aggregation, where reversible means that from any state of our
system we can recover the previous state exactly. We must address the
subtleties of making each component step invertible, including steps
which realize stochastic processes.  As we discuss below, the same
mechanisms that are employed in our model in order to ensure exact
conservation of energy, particle number, and other constraints, also
make it easy to incorporate invertibility.  
The stochastic component of the model is diffusion,
which is modeled as a sequence of invertible ``random walks'' based on
a deterministic algorithm using an invertible pseudorandom number
generator.

Information about exactly when and where a particle undergoes a phase
transition is stored in the heat particles.  The
idea of storing information in a heat bath was introduced by
Creutz\cite{Creutz-annphys86} to explore the connection between the
microcanonical and canonical ensembles for the dynamical Ising
model. Heat bath techniques have been used on occasion since then to
construct reversible computer models of physical
phenomena\cite{Smith-phd}.

\subsubsection{Overview of the model} \label{ra-gen-descrip}

To construct the reversible aggregation (RA) model 
we begin with a parallel DLA system (similar to
that described above) and add degrees of freedom at each site,
corresponding to a distributed heat bath. The latent heat released
during each aggregation event can then be explicitly represented.  In
the RA dynamics, whenever a random walking gas particle lands on a site
with exactly one nearest neighbor crystal particle, it will stick only
if there is room locally in the heat bath to accept the latent heat it
will release as the particle transitions from the gas phase to the
crystal phase.  The heat is released in quantized units called heat
particles, with one heat particle released for each aggregation event.
These heat particles diffuse amongst themselves  (i.e., they undergo
random walks along the lattice sites, independently of the gas
particles). Explicitly modeling the latent heat released upon 
aggregating provides a mechanism for modeling the inverse process: a
diffusing heat particle which contacts a susceptible crystal particle (a
crystal particle which has only one nearest neighbor crystal particle)
is absorbed while the crystal particle evaporates to become a
gas particle which then diffuses away. 

The restriction on the dynamics that aggregation and evaporation
events can occur only when exactly one nearest neighbor is a crystal
particle means that only one crystal bond is ever formed or broken
when a single lattice site is updated.  As each aggregate particle
contributes one crystal bond to the aggregate, and there is no
further potential energy contribution, the energy of the
aggregate is a function only of the number of aggregate particles,
independent of their configuration.  Moreover this constraint has two
direct implications for the growth morphology.  The first is that
evaporation can only occur for particles which are singly connected to
the growth cluster, and so the aggregate cannot break off into
disconnected clusters.  The second is that it introduces an excluded
volume (i.e., no closed loops can be formed), thus we might expect the
equilibrium cluster configuration to be similar to that of a polymer.
Note that introduction of an evaporation mechanism in the RA model
mitigates the shadowing effect that was important in determining the
DLA growth morphology: crystal particles within the cluster can
evaporate, thus introducing gas particles into the interior of the
aggregate.

\subsubsection{The detailed dynamics}\label{sec.7bits}

The RA model is constructed with 7-bits of state at each lattice site.
One bit, $N_c(\x,t)$, denotes the presence or absence of a crystal
particle at that site (i.e., $N_c(\x,t)=1$ indicates presence,
$N_c(\x,t)=0$ absence).  Two bits, $N_g^{\gamma}(\x,t)$ where $\gamma =
\{1,2\}$, denote the presence or absence of each of two gas particles.
Two bits, $N_h^{\gamma}(\x,t)$, denote the presence or absence  
of each of two heat particles. The final 2 bits, $\xi_g(\x,t),
\xi_h(\x,t)$, are independent binary pseudorandom variables. The
dynamics of the model consists of two kinds of steps: diffusion steps
alternating with interaction steps.

The same kind of diffusion process is applied to the gas and heat
subsystems simultaneously and independently, while the crystal
particles remain unchanged.  A given diffusion step consists of 
two parts: mixing and transport.  During the mixing portion of the
step, a binary random variable is used to determine whether or not the 
two particle bits of that species at the site $(\x,t)$ are
interchanged: 
\begin{eqnarray}
N^1_i & = & (1-\xi_i)N^1_i + \xi_i N^2_i
\nonumber \\ 
N^2_i & = & (1-\xi_i)N^2_i + \xi_i N^1_i,
\end{eqnarray}
where $i=g$ or $i=h$.
During the transport portion of the step, every site replaces its
first particle bit ($\gamma=1$) with the first particle of its
neighbor a distance $k$ away on one side, and its second
particle ($\gamma =2$) with the one from the same distance neighbor on
the opposite side.  At even time steps, we  
use horizontal neighbors (i.e., the diffusion moves particles
horizontally):
\begin{eqnarray}\label{eq.trans-horz}
N^1_i(\x,t+1) & = & N^1_i(\x+k\hat{x},t) 
\nonumber\\
N^2_i(\x,t+1) & = & N^2_i(\x-k\hat{x},t).
\end{eqnarray}
At odd time steps we use vertical neighbors (i.e., substitute
$\hat{y}$ for $\hat{x}$ in Eq.~(\ref{eq.trans-horz})). The
only differences between the gas and heat diffusion are (1) each uses
a separate binary random variable to control its mixing, and (2) the
distance of the neighbor particle to be copied,
$k$, can be chosen separately for each subsystem---this
allows us to independently control the diffusion constants for the
heat bath and for the gas (c.f. \cite{ChopDroz-diff}).

Diffusion steps alternate with steps in which the two
subsystems---gas-crystal and heat bath---interact allowing
aggregation and evaporation.  The rule at a single lattice site during
an interaction step is that exactly one particle can aggregate or
evaporate provided that
 \begin{description}
 \item[(a)] there is exactly one crystal particle at one of the four
nearest neighbor sites 
 \item[(b)] there is room at the site for a crystal particle (for
aggregation) or for another gas particle (for evaporation), and
 \item[(c)] the heat bits at the site can absorb (for aggregation)
or supply (for evaporation) one unit of heat.
 \end{description}
Since the gas and heat particles will undergo a mixing step before
transport, it makes no difference which of the two available gas
particle positions a crystal particle is moved into when it evaporates,
or which of the two possible heat particle positions a unit of heat
gets put into.  Defining this precisely will, however, become
important when we discuss invertibility.

The interaction rule described thus far would be sufficient if we
updated just one lattice site at a time. If, however, all sites on the
lattice are updated simultaneously, then the global dynamics no longer
obeys the ``single bond'' constraint---that at any site where
particles aggregate or 
evaporate exactly one crystal bond is formed or broken. For example,
suppose that the tip of a crystal branch evaporates just as a gas
particle condenses next to it. Each of these events would separately
preserve the constraint, but the two simultaneous events result in the
addition of a disconnected crystal particle which has no other 
crystal particle immediately adjacent to it.  We can easily avoid this
difficulty by holding the values at the adjacent sites fixed during a
step in which we let the subsystems interact at a given lattice
site, since the interaction step has a nearest neighbor range. In other 
words, we perform a {\em checkerboard updating}: all of the lattice
sites in which the $x$ and $y$ lattice coordinates add up to an even
number are updated using our single site interaction rule, while the
odd sites are held fixed, and then vice versa.  Since nearest neighbors
are held fixed during an interaction, the constraint is obeyed locally
and thus it is also obeyed globally.  The overall dynamical
rule is summarized in Table~\ref{tab.rule}---the various phases 
of the rule are applied consecutively.

\begin{table}[tbph]
\centerline{\fbox{\begin{minipage}{3in}{\small\tt
\begin{enumerate}
 \item interact gas/heat/crystal at even sites
 \item mix gas and mix heat separately
 \item transport gas and heat horizontally
 \item interact gas/heat/crystal at odd sites
 \item mix gas and mix heat separately
 \item transport gas and heat vertically
\end{enumerate}}\end{minipage}}}
\medskip\noindent\caption{The various phases of one step of the RA
dynamics.  Each phase is applied over the entire lattice
simultaneously.} 
\label{tab.rule}
\end{table}

Every phase of the rule described in Table~\ref{tab.rule}
can be inverted. The transport portion of the step can be run
backwards by simply moving all particles back into the sites they came
from (i.e., inverting the directions of the transport in
Eq.~\ref{eq.trans-horz}).   The mixing portion of the step is easy 
to invert, given the same ``random'' binary variables that
were used to determine which pairs of bits were originally swapped.  We 
simply swap exactly those pairs again.  The pseudorandom portion of
the system (which supplies the random data) can simply follow some
invertible dynamics that is independent of the rest of the
system---the rest of the system looks at the state of this subsystem
but does not affect it---so this pseudorandom subsystem can be run {\em
backwards} independently of everything else. 

Making the interaction steps invertible is also straightforward.  When
a single gas particle turns into a crystal particle, we
put the heat token that is released into the corresponding heat bit
(i.e., the heat particle with the same value of $\gamma$), and
thus we remember which of the two gas particle bits was initially
occupied.  If the corresponding heat bit is already occupied, the
particle is not allowed to aggregate (even if the other heat bit is
unoccupied). Similarly, a crystal particle is allowed to evaporate only
if it can move into the gas bit with the same value of $\gamma$ as the
heat token being absorbed.  If there are two gas particles at a
site we impose the constraint that the particle with $\gamma=1$
attempts to aggregate first; likewise if there are two heat particles 
at a site, the crystal particle attempts to absorb the $\gamma=1$
particle first. This does not introduce a bias to the growth since we
are mixing the $\gamma=1$ and $\gamma=2$ variables in an unbiased
manner at each time step.  
With these refinements, our interaction rule {\em 
applied to a single site} is its own inverse: if we apply it twice
(without a diffusion step in between) we get back the state we started 
with.  Since the interaction rule is applied in a checkerboard
fashion, sites are updated independently: if we apply the rule a
second time to the same checkerboard, it will undo the first
application at every site. 

Thus an inverse step consists of
applying the inverses of the rule-phases in the opposite of the order
listed---once one phase is undone, the previous phase can be undone.
Each inverse step undoes one step of the forward time evolution.
As we watch the inverse evolution, we see each heat particle retrace
its path, to be in exactly the right location at the right time to
uncrystalize the crystal particle which originally released it.
Particles un-aggregate and un-diffuse and un-evaporate in a manner
that exactly retraces their behavior in the forward evolution.

\subsubsection{Implementation} \label{implement}

The RA model was implemented on a special purpose cellular automata
machine, the CAM-8\cite{cam8-waterloo}, which was designed to
efficiently perform large-scale uniform, spatially arrayed
computations.  On this machine, all simulations must be embedded into
a lattice gas framework\cite{HPP-pra76,FHP-prl86}, in which uniform
data movement (data-advection) alternates with processing each site
independently (site-update).  For a 2-dimensional model such as ours,
sheets of bits move coherently during the advection phase:
corresponding bits at each site all move in the same direction by the
same amount.  The boundaries are periodic---bits that shift past the
edge of the lattice reappear at the opposite edge.  After moving the
bits, we perform the site update phase.  During this phase, the bits
that have landed at each lattice site are updated in a single
operation by table lookup: the bits at each lattice site are used as
an index into a table that contains a complete listing of which state
should replace each possible original state.  Both the data movement
and the lookup table can be freely changed between one
lattice updating step and the next.

Our model requires 7 bits of state to appear at each site
in our $L\times L$ lattice.  Using random data generated by a serial
computer, the bits which correspond to the gas
particles are initially randomly filled with a $4\%$ density of
particles and the bits which correspond to the binary random
variables with a $50\%$ density of particles.   One crystal particle
is placed at the center site of the lattice. The heat bath is
initially empty. 

The dynamics on the pseudorandom subsystem is very simple:
each of the two random bit-planes (each consisting of all the
$\eta_g$'s or all of the $\eta_h$'s) are simply shifted by some large
amount at  each time step.  We could choose the amount and direction
of each shift  at random for each step of updating, using a reversible
random number generator running on the workstation controlling the
simulation. Instead, the simulations discussed here simply shift the
bit planes by a large and fixed amount at each step, making sure that
the the $x$ and $y$ components of the two shifts are all mutually
co-prime, as well as being co-prime with the overall dimensions of the
lattice. Thus to run the random subsystem backwards, we just reverse
the direction of the shifts.

The checkerboard updating is accommodated by adding an eighth bit to
each lattice site, and filling these bits with a checkerboard pattern
of ones and zeros.  In our rule the various subsystems are allowed to
interact only at sites marked with a one.  To change which
checkerboard is marked for updating, we simply shift the
checkerboard-marker bit-plane by one position in the $+x$ direction.

The rule described in Table~\ref{tab.rule} turns into two lattice-gas
steps on CAM-8. The first three phases listed in Table~\ref{tab.rule}
are done during one step, and the next three in the second step.  The
data movement is part of each step: note that each of the two steps
uses the same lookup table applied to each lattice site, but slightly
different data movement.  To run backwards, we use the inverse lookup
table, and the inverse data movement. Note that in the discussion of
experimental results, everything in Table~\ref{tab.rule} is counted as
a single step. 

CAM-8's event counting hardware was used to monitor simulation
parameters while the simulations ran.  Including event counting, the
8-processor CAM-8 performed about $10^8$ site update operations per
second for this model.

\section{The macroscopic limit} \label{analytics}

The dynamics of the RA model, described in detail above, can be
succinctly presented in an analytic framework.  We develop this
framework first in terms of the discrete space, time, and occupation
number variables.  We then ensemble average over the occupation
numbers and take the continuum limit of the space and time
variables to establish the reaction-diffusion equations for the
system.   

As discussed in Sec.~\ref{sec.7bits} there are seven bits of state:
$N_c(\x,t), N_g^{\gamma}(\x,t), N_h^{\gamma}(x,t), \xi_g(x,t),$ and
$\xi_h(x,t)$, where $\gamma=\{1,2\}$. They correspond respectively to 
one bit of crystal, two bits of gas particles, two bits of heat
particles, and two bits of random data. 
$N_i^{\gamma}(\vec{x},t)=1$ indicates the presence
of species $i$ at location $\vec{x}$ and time $t$, in channel
$\gamma$, and $N_i^\gamma(\vec{x},t)=0$ indicates the absence. 
The absence or presence of a crystal particle is denoted by
$N_c(\vec{x},t) = \{0,1\}$ respectively. The
total number of particles of species $i$ at time $t$  
present on the lattice is denoted by $\CN_i(t) = \sum_{\x, \gamma}
N_i^{\gamma}(\x,t)$, where the sum is over all of the lattice sites
and the two particle channels.

There are no external sources or sinks for any of the three species
represented (the gas, crystal, and heat species). Energy is only
exchanged between the gas-crystal and the heat bath subsystems.  Thus 
the complete system is isolated. Conservation of the total number of
gas and crystal particles implies that 
\begin{equation}
\CN_g(t) + \CN_c(t) = \CN_g(0) + \CN_c(0). \label{mass-cons}
\end{equation}
Conservation of the total energy of the system implies that
\begin{eqnarray}
\CN_g(t) \varepsilon_g + \CN_c(t) \varepsilon_c + & \CN_h(t)
\varepsilon_h & =  \nonumber \\
& \CN_g(0) \varepsilon_g & + \CN_c(0) \varepsilon_c + \CN_h(0)
\varepsilon_h, \label{energy-cons} 
\end{eqnarray}
where $\varepsilon_i$ represents the energy (kinetic and
potential) per particle of species $i$ (notice that there is no
configurational contribution to the energy of the crystal).  As discussed in
Sec. \ref{sec.mod-agg} each aggregation event releases one heat particle
(likewise each evaporation event absorbs one heat particle), thus
$\varepsilon_h = \varepsilon_g - \varepsilon_c$ and moreover 
\begin{equation}
\CN_h(t) = \CN_c(t) - \CN_c(0). \label{heat-cryst-cons} 
\end{equation}
(Note that $\CN_h(0) = 0$ and $\CN_c(0) = 1$ in our experiments).

To facilitate the description of the dynamics, we introduce a
functional equation which is $+1$ at any site where a particle is
about to crystalize, $-1$ at a site where a particle is about to
evaporate, and $0$ otherwise.  The functional,
$\CF^\gamma\left[N_g^\gamma(\x,t), N_h^\gamma(\x,t), N_c(\x,t), 
\left\{N_c(\x+\e_k,t)\right\} \right]$, is evaluated on a neighborhood
of lattice sites surrounding some given position $\x$ at a given time
$t$ (the notation $\left\{N_c(\x+\e_k,t)\right\}$ refers to the set
of values of $N_c$ for the nearest neighbors of the point $\x$).

\end{multicols}\setlength{\columnwidth}{\hsize}
\ \ \ \ \ \ \ \ \hrulefill \ \ \ \ \ \ \ \ 
\begin{eqnarray}
\CF^\gamma(\x,t) = N_g^\gamma(\x,t) [1-N_h^\gamma(\x,t)] [1-N_c(\x,t)]
\sum_{j=1}^d N_c(\x + \e_{j},t) 
\prod_{k\neq j} \left[1 - N_c(\x+\e_k,t)\right] 
\nonumber \\
- [1-N_g^\gamma(\x,t)] N_h^\gamma(\x,t) N_c(\x,t) 
\sum_{j=1}^d N_c(\x + \e_{j},t) 
\prod_{k\neq j} \left[1 - N_c(\x+\e_k,t)\right]. \label{F}
\end{eqnarray}
\begin{multicols}{2}\setlength{\columnwidth}{\hsize}
\noindent
Here $\e_{j}$ and $\e_{k}$ are the vector lattice
directions of the nearest neighbor cells, and $d$ the number of
distinct lattice directions. For a two dimensional square lattice
(i.e., the lattice used for the present implementation) $d=4$ and
the vector lattice directions are $\left\{\hat{x}, -\hat{x},
\hat{y},-\hat{y}\right\}$. 

The first term in Eq. (\ref{F}) equals $1$
if a gas particle in channel $\gamma$ is present at site $\x$
and time $t$, a heat particle in channel $\gamma$ is absent
at site $\x$ and time $t$, there is no crystal particle already at
that site, and only one crystal particle is present at a nearest
neighbor site. It is zero otherwise. The second term equals $1$ if 
there is no gas particle in channel $\gamma$ present at site
$\x$ and time $t$, there is a heat particle
in channel $\gamma$ present at site $\x$ and time $t$,
there is a crystal particle present at that site,   and only one crystal
particle is present at a nearest neighbor site. It is zero otherwise.
The first and second terms are mutually exclusive (a heat particle
in channel $\gamma$ cannot be simultaneously present and
absent, nor can a gas particle).

The dynamics consists of making the changes indicated by $\CF^1$
and then $\CF^2$, then applying a random permutation to mix $\gamma=1$ and
$\gamma=2$, and then performing the streaming step to move the
particles.  The permutation mixes the $N_i^1(\x,t)$ and $N_i^2(\x,t)$ 
components in an unbiased way, so it is simpler to discuss the
dynamics of a combined variable, $N_i(\x,t)=N_i^1(\x,t)+N_i^2(\x,t)$.
Likewise, if we let 
$\CF(\x,t)=[\CF^1(\x,t) + \CF^2(\x,t)][1-\CF^1(\x,t)\CF^2(\x,t)/2]$,
the interaction portion of the dynamics at a single lattice
site can be written 
\begin{eqnarray}\label{eq.interaction} 
N_c(\x,t+1) & = & N_c(\x,t) + \CF(\x,t) \nonumber \\
N_g(\x,t+1) & = & N_g(\x,t) - \CF(\x,t) \\
N_h(\x,t+1) & = & N_h(\x,t) + \CF(\x,t) \nonumber 
\end{eqnarray}
The $[1-\CF^1(\x,t)\CF^2(\x,t)/2]$ factor in the definition of
$\CF(\x,t)$ reflects the fact that only one 
transition occurs at a given site, even if two gas or two heat
particles are present. 

To construct continuous variables from the discrete ones, we consider
the average short-term behavior of the system over an ensemble of
independent realizations which all have the same set of local particle
densities.  For each discrete variable, we let $n_i(\x,t)\equiv
\left<N_i(\vec{x},t)\right>$, and denote the average of the 
functional as $\left<\CF\right>$.  (This technique of
averaging over many independent realizations, i.e., establishing the
one particle density function,  is commonly used to
derive the lattice Boltzmann equation starting from discrete particle
models of hydrodynamics\cite{Bmb-renorm,Hassl-discretefluids}). 

With this notation, the average
propagation of the gas and heat particles can be expressed as
\begin{eqnarray}
 n_i(\x,t+1) = \nonumber \\
 \frac{1}{4} \left[n_i(\x-\hat{x},t) + n_i(\x-\hat{y},t) +
 \right. \nonumber \\
 \left. + n_i(\x+\hat{x},t) + n_i(\x+\hat{y},t) \right].
\end{eqnarray}
To establish the continuum limit, we Taylor expand. The terms
involving the first derivatives cancel, leading to the result
 \begin{eqnarray}\label{eq.diffusion} 
 n_i(\x,t+1) = n_i(\x,t)  + \nonumber \\
 \sum_{j} \frac{|\Delta x|^2}{4} \partd{^2}{x_{j}^2} 
\left(n_i(\x,t) \right) + O(\Delta x^3)   \nonumber \\
 = n_i(\x,t) + \frac{|\Delta x|^2}{4} \nabla^2
n_i(\x,t),
\end{eqnarray}
where $i=g$ or $i=h$. Note that to order $\Delta t$ the above equation
is the standard diffusion equation,
 \begin{equation}\label{eq.simp.diff}
 \partd{}{t} n_i(\x,t) = \frac{|\Delta x|^2}{4\Delta t} \nabla^2
 n_i(\x,t).
 \end{equation}

As discussed in Sec.~\ref{sec.7bits}, we can control the length of
each diffusion step separately for the heat and for the gas
particles.  The heat particles execute
random walks composed of individual steps of length
$k$, whereas the gas particles execute walks of step length 
unity.  Thus if the $|\Delta x|^2$ that appears in
Eq.~(\ref{eq.diffusion}) and in Eq.~(\ref{eq.simp.diff})
refers to the gas subsystem, then $|\Delta
x|^2$ for the heat subsystem (and hence its diffusion constant) is a
factor of $k^2$ larger.

To proceed further, we will make the approximation that our average
variables are independent.  Then we are allowed to replace the average
of a product of variables by the product of the average for each
variable: $\left<ab\right> = \left<a\right>\left<b\right>$, for $a,b$
independent variables.  This is the assumption of molecular chaos,
which is also used to derive the lattice Boltzmann equation. 
With this approximation the average of the functional $\CF^\gamma$
is simply 
\begin{eqnarray}
& \left<\CF^\gamma\right> = 
\left\{n_g^\gamma(\x,t)\left[1-n_h^\gamma(\x,t)\right]\left[1-n_c(\x,t)\right]\right.
\nonumber \\
& \left. - \left[1-n_g^\gamma(\x,t)\right]n_h^\gamma(\x,t)n_c(\x,t)\right\}  \nonumber \\
& \times \sum_j n_c(\x+\e_j,t)\prod_{k\neq j} \left[1-n_c(\x+\e_k,t)\right].
\label{eq.fgamma}
\end{eqnarray}
Similarly, we can write down an expression for $\left<\CF\right>$.  

To obtain the continuum limit of these averaged equations, we again use
Taylor series approximations.  In the diffusive regime, 
$\Delta t \sim \left(\Delta l \right)^2$, so we truncate the expansions 
at these appropriate orders.  Let $\tilde{\CF}$ be the
continuum limit of $\left<\CF\right>$ (which we won't write out
explicitly).  Then from Eq.~(\ref{eq.interaction}), we obtain
  \begin{eqnarray}
  \partd{}{t} n_c(\x,t) & = & \frac{1}{\Delta t} \tilde{\CF}, \label{eq.react-c}
  \end{eqnarray}

The other reaction-diffusion equations for our system can be obtained
by proceeding as we did in Eq.~(\ref{eq.diffusion}).  For example,
under the full dynamics (which consists of both the interaction and
diffusion phases)  
  \begin{eqnarray}
  & n_g(\x,t+1) + n_c(\x,t+1) = n_c(\x,t) + \nonumber \\
  & \frac{1}{4} \left[n_g(\x-\hat{x},t) + n_g(\x-\hat{y},t) +
\right. \nonumber \\
  & \left. n_g(\x+\hat{x},t) + n_g(\x+\hat{y},t) \right],
  \end{eqnarray}
since any particles present at a site at time $t+1$ were either
already there at time $t$, or moved there.  Expanding this exactly
as in Eq.~(\ref{eq.diffusion}) and using Eq.~(\ref{eq.react-c}), we get
  \begin{eqnarray}\label{rdg}
  \partd{}{t} n_g(\x,t) & = & \frac{|\Delta x|^2}{4 \Delta t} \nabla^2
  n_g(\x,t) - \partd{}{t} n_c(\x,t), 
  \end{eqnarray}
and similarly, 
  \begin{eqnarray}\label{rdh}
  \partd{}{t} n_h(\x,t) & = & \frac{k^2 |\Delta x|^2}{4 \Delta t}
  \nabla^2 n_h(\x,t) + \partd{}{t} n_c(\x,t).
  \end{eqnarray}

Note that if $k=1$, and we add the last two equations 
the variable $n_g(\x,t)+n_h(\x,t)$ obeys the diffusion equation,
unaffected by the interaction between the subsystems (i.e. if we
remove the distinctions between gas and heat, the combined variable
simply diffuses without interacting). 

To test the consistency between the microscopic diffusive dynamics
implemented in our model and macroscopic descriptions given by
Eq.~(\ref{rdg}) and Eq.~(\ref{rdh}), we empirically measured the
diffusion coefficient for gas and for heat particles as they diffuse 
about the space. Each particle should execute a random walk. The
variance of the distance from the origin in the $\hat{x}$ or $\hat{y}$
direction, $\sigma^2_i$, is proportional to the diffusion coefficient
in that direction, $D_i$, where $i=g$ or $i=h$.  The exact relation
is $D_i = \sigma^2_i/4p$, where $p$ is the number of steps taken. For
an unbiased random walk, the variance of the net displacement from the
origin is $\sigma^2_i = k^2 p$, thus $D_i = k^2/4$.
For the gas particles ($k=1$) we find $D_g = (0.996 \pm 0.009)/4$.
For the heat particles, with $k=3$, we find $D_h = (9.00 \pm 0.08)/4$.   
Thus the ratio of the heat to the gas diffusion length 
$D_h/D_g = 9.0\pm 0.1$, agreeing with the theoretically predicted
value of $k^2$.

\section{The mean field limit} \label{meanfield}

The mean field limit corresponds to the ``well-stirred reaction,''
meaning that the reacting species are uniformly spread throughout the
space, and thus each particle feels the presence of the mean
concentration of each species.  For our system in equilibrium the gas
particles and the heat particles {\em are} uniformly distributed
throughout the space; it is only the crystal particles which do not
obey this assumption. A uniform distribution means that there are no
concentration gradients ($\nabla n_i(\x,t) = 0$ for all $\x$ and $i$),
and thus $\nabla^2 n_i(\x,t) = 0$. Also we can drop the explicit
$\vec{x}$ notation from the argument of the variables: $n_i(\x,t) =
n_i(t)$. 
  
Once the population level of the heat bath has reached the quasistatic
steady-state, the concentrations of all three species will 
remain essentially constant and the systems will have a
well defined temperature from then on, as discussed in
Sec.~\ref{temp}.  We denote the time to reach the steady-state
(i.e. the time for the subsystems to reach the same temperature) as
$\tau_{\rm{T}}$. We can now drop the explicit time notation from the
arguments of the variables in steady-state:  $n_i(t > \tau_{\rm{T}}) =
n_i$.   In this regime $dn_i/dt = 0$ and likewise $dn_i^\gamma/dt =
0$, thus $<\CF^\gamma> = 0$: 
\begin{eqnarray} \label{mf-ss}
0 & = & \left<\CF^\gamma\right> \nonumber \\
  & = &
 4 \left[
n_g\left(1-n_h\right)\left(1-n_c\right) - \left(1-n_g\right) n_h n_c 
\right] \times \nonumber \\
  & & \left[n_c \left(1-n_c\right)^3\right].  
\end{eqnarray} 

\begin{figure}[tbph]
\hfill\vbox{%
\psfig{figure=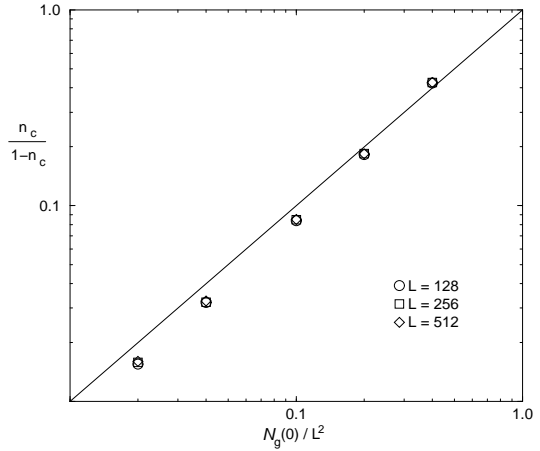,width=2.5in}%
}\hfill
\medskip\caption{The empirically determined equilibrium value of
$n_c/(1-n_c)$ as a function of the initial gas density, $\CN_g(0)/L^2$,
for systems of size $L=128,\ 256,{\rm \ and \ } 512$.  The solid line
is the mean field prediction. Note that the error bars are
the same size as the points.}
\label{fig1.mf}
\end{figure}

There are three solutions to Eq. (\ref{mf-ss}). Each solution corresponds
to fixed point of the dynamics. Only one 
is in the regime of interest. The fixed point at $n_c=0$ corresponds
to the presence of only gas particles.  The fixed point at $n_c=1$ is
not allowed by the aggregation conditions (the aggregate can not have
any closed loops).  The remaining fixed point predicts that the 
equilibrium condition is  $n_g \left[1-n_h\right]\left[1-n_c\right]
 = \left[1-n_g\right] n_h n_c.$
Noting that in the mean field limit ensemble averages 
equal spatial averages (i.e., $n_g(t) =  \CN_g(t)/L^2$), 
the constraints described in Eq. (\ref{mass-cons}) and
Eq. (\ref{heat-cryst-cons}) can be written respectively as
$n_g(t) = \CN_g(0)/L^2 - n_c(t) + 1/L^2 \approx \CN_g(0)/L^2 - n_c(t),$ 
and $n_c(t) = n_h(t) + 1/L^2
\approx n_h$. 
After incorporating these relations the equilibrium condition can be
expressed as 
\begin{equation}
\frac{n_c}{1-n_c} = \CN_g(0) \frac{1}{L^2} + O(n_c^3). \label{nc_eqm}
\end{equation}
Figure \ref{fig1.mf} is a plot of the equilibrium value $n_c/(1-n_c)$
versus the initial density of gas particles, $\CN_g(0)/L^2$. The solid
line is the mean field prediction, Eq. (\ref{nc_eqm}). The points were
obtained empirically from our simulations of three different system  
sizes, $L=128,\ 256, {\rm \ and \ } 512$.  The agreement between the
three system sizes should be noted. 

\bigskip

The mean field approach makes predictions about the overall density of
the system (hence the equilibrium temperature, as described below),
but it does not make predictions about the growth morphology.

\section{Empirical analysis}

\subsection{Temperature} \label{temp}

The gas-aggregate subsystem and the heat bath subsystem together form
a thermodynamically isolated system.  These two subsystems are allowed
to exchange energy only between themselves, and this energy is purely
in the form of heat ($\Delta Q$). As discussed in
Sec.~\ref{ra-gen-descrip} and Sec.~\ref{analytics} the energy of the
aggregate is a function only of the number of aggregate particles and
is independent of their configuration. The total internal
energy of the gas and heat particle species 
also is a function only of the number of particles of each
species. Hence if invariant average population densities are achieved
there is no further net exchange of heat between the subsystems,
and they have by definition attained the same temperature.

The standard expression for the temperature of a two-level
system\cite{Huang-statmech}, such as the heat bath in the RA model,
follows directly from combining the definition of temperature ($1/T =
\left. \Delta S/ \Delta E \right|_{V}$) with the microcanonical
definition of entropy ($\Delta S=k_B \ln(\Omega_f/\Omega_i)$, where
$\Omega$ denotes the number of microstates consistent with the
macroscopic variables):
  \begin{equation}
  \frac{1}{T} = -\frac{k_B}{\varepsilon_h}
  \ln\left(\frac{n_h}{1-n_h}\right). 
  \end{equation}
Directly computing the temperature of the other subsystem is more
difficult, but we can infer its temperature from that of the heat
bath (note the gas particles are free to diffuse over the crystal,
hence there is no change in the accessible volume for the heat
particles as the crystal changes size and conformation: the crystal
does no work on the gas, $PdV=0$). 

\begin{figure}[tbph]
\hfill\psfig{figure=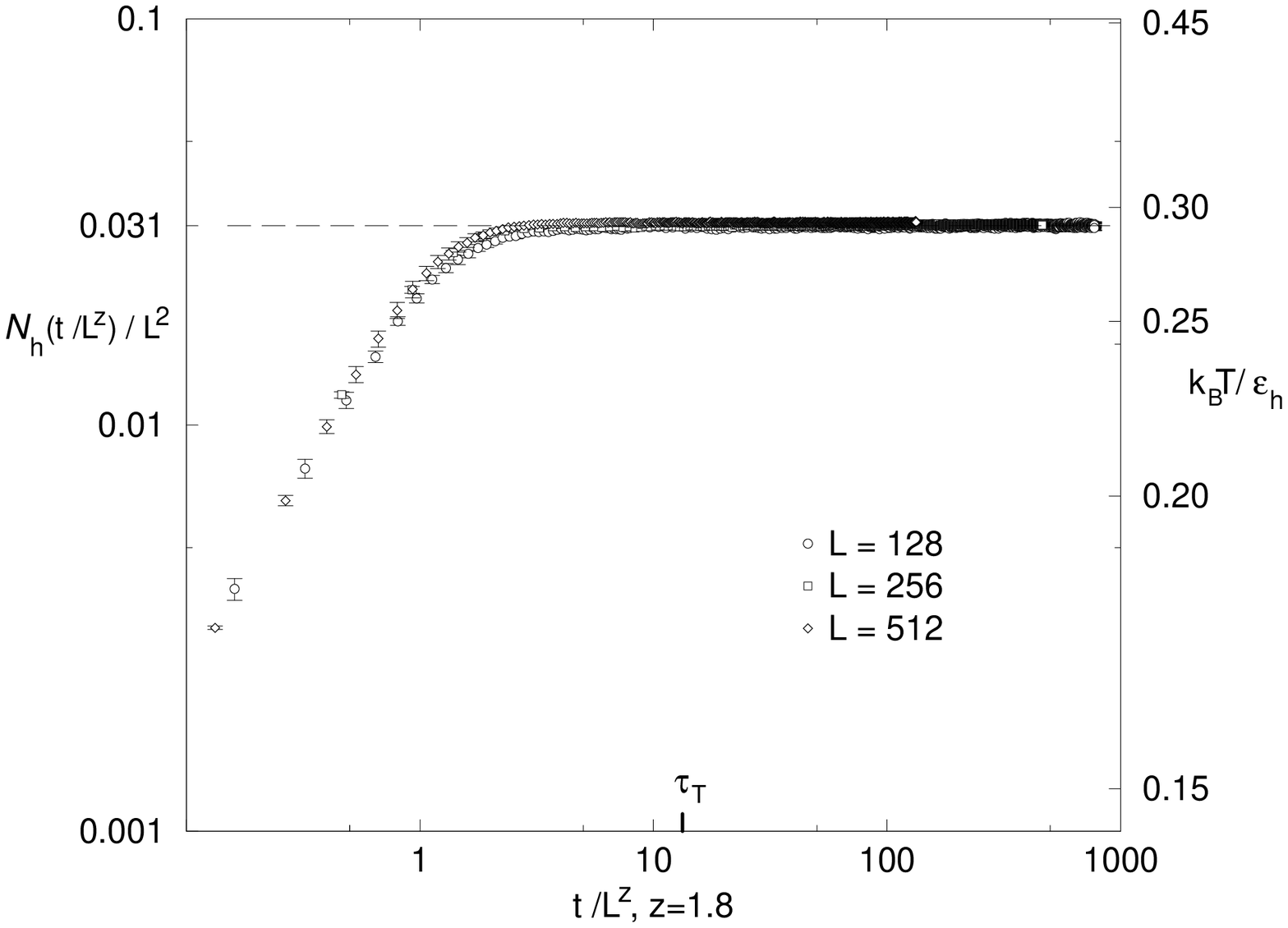,width=3in}\hfill
\psfig{figure=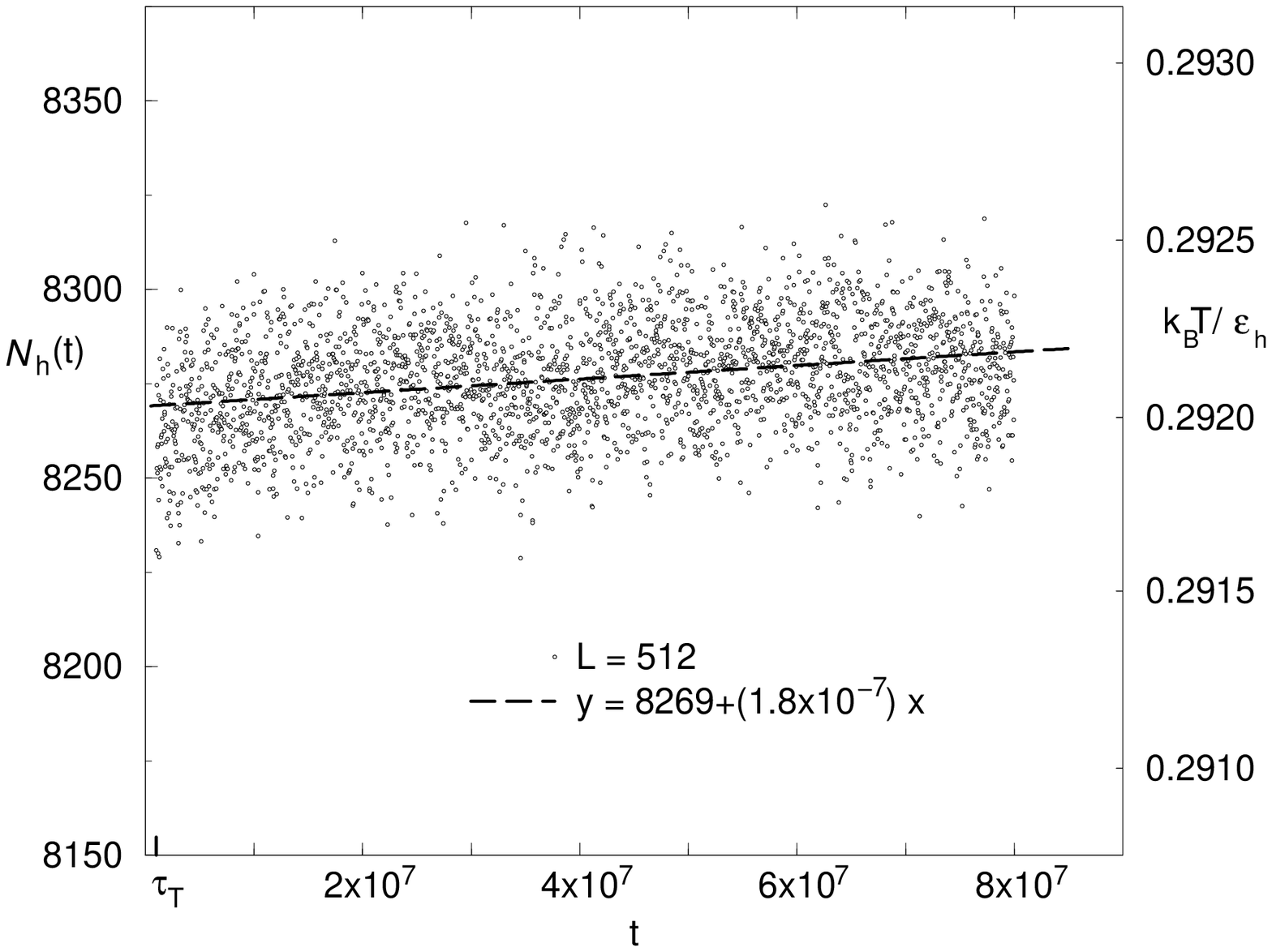,width=3in}\hfill
\medskip\caption{(a)~The mean density of heat bath particles as a
function of time into the simulation, plotted for $L=128,\  256, {\rm
\ and \ } 512$. The corresponding temperature is given on the right
vertical axis. The initial growth of the heat bath
density is linear, with a slope of about 1.8. Note that the
steady-state density of the heat bath, and hence the temperature, is
equal for all three system sizes, yet the time to equilibrate scales
with the system size as $\tau_{\rm{T}} \sim 10 L^{1.8}$.   
(b)~The actual average values of the total population of the heat bath
as a function of time, for every third time measured beyond
$\tau_{\rm{T}}$. The dashed line is the result of a linear least
squares regression on all of the data. Note the slight drift upward
with time, of about 2 particles per $10^7$ steps.}  
\label{fig2.temp-eqm}
\end{figure}

The approach to temperature equilibrium and a closeup of the
subsequent fluctuations in temperature are shown in Fig.~\ref{fig2.temp-eqm}. 
Figure~\ref{fig2.temp-eqm}(a) plots the mean occupancy of the heat bath
versus the time step into the simulation, with the corresponding
temperature (in units of $k_B T/\varepsilon_h$) displayed on the right
vertical axis.  The initial growth is linear, with a slope of about
1.8. It then levels off near the  quasistatic steady-state density of
$n_h = 0.031$ (indicated in the figure as the dashed horizontal line).
The results are averages over several independent realizations for 
three different system sizes, $L=128,\  256, {\rm \ and \ } 512$.  
Note that the three systems reach the same steady-state densities and
hence the same temperature, but that the time to equilibrate depends
on the system size.   The data for the three systems was collapsed
onto one curve by rescaling the time axis by the factor $L^z$, with
$z=1.8$. The time to reach the equilibrium temperature is
$\tau_{\rm{T}} \sim 10 L^z$. Note that this scaling behavior has an
exponent which is slightly smaller than the diffusion exponent: the
diffusion time is proportional to $L^2$. As discussed in
Sec.~\ref{frac-dim-results} the fractal dimension at time
$\tau_{\rm{T}} \approx 1.8$: the time seems to scale with the fractal
dimensionality instead of the Euclidean dimensionality of the space. 

To study the details of the subsequent fluctuations we focus on
the largest system size, $512\times512$.  As mentioned,
during the initial period the growth of the heat bath population (and
the size of the aggregate) is linear. It levels off at about
8270 particles on average, in a time which is less than $10^6$ steps.  
The population continues to grow extremely slowly after this 
point, rising by an average of $1.8\pm0.1$ particles every $1\times
10^7$ steps, as determined by a linear regression based on about 8000
data points taken at equally spaced intervals in the regime where
$t > \tau_{\rm{T}}$.  The probability that the slope is actually zero is  
$3\times 10^{-8}$, as determined by a t-statistic comparing the 
ratio of the obtained slope to the sum of squares differences. 
Figure~\ref{fig2.temp-eqm}(b) shows a scatter plot of every third
of the 8000 data points, overlayed by a straight line indicating
the results of the linear regression on all 8000 points (only every
third point is shown for visualization purposes: showing all
the points results in a dense black cloud). The  actual number of
particles in the heat bath is indicated on the left axis, the
corresponding temperature is given on the right.  Although the
temperature of the heat bath is not constant, it is very nearly
so. Once the population levels stabilize the subsequent dynamics
(i.e., the relaxation to the maximum entropy state for the crystal) is
clearly a quasistatic process.  The crystal does continue to exchange
heat with the heat bath when it anneals, but the net heat exchange 
is essentially zero (the net heat exchange rate is $\sim 2\times 10^{-7}$
particles per update of the space). 

We measured also the fluctuations in population levels of
the heat bath for the $L=256$ and $128$ systems, in the corresponding
regimes. Consistent with the $L=512$ system, we find the population 
rises by an average of $2.1\pm0.1$ particles every
$1\times 10^7$ steps.  

\end{multicols}\setlength{\columnwidth}{\hsize}

\bigskip
\begin{multicols}{2}\setlength{\columnwidth}{\hsize}

\subsection{Fractal Dimension} \label{frac-dim-results}

The aggregate formed primarily while the heat bath contained less
energy than its equilibrium level. Hence, if we continue running the
dynamics, the cluster anneals; it evolves from a DLA-like cluster to a
quenched branched random polymer structure.  To quantify the cluster
structure we calculate the fractal dimension of the aggregate and
specifically how the fractal dimension changes as a function of
the time into the simulation. 

We measure the fractal dimension using a box-counting procedure which
requires that we first establish the center of mass of the growth
aggregate (which is typically not the initial seed particle, as the
center of mass diffuses about the space as the aggregate anneals).
An imaginary window box of edge length $l$ is defined and
centered on the center of mass. The number of lattice 
sites within that window that contain a crystal particle, $\CN_c(l)$,
is tallied.   The window size is increased and the count retallied.
This procedure is iterated until the number of crystal
particles contained no longer increases with window
size.   Before saturation, the number of particles contained should
increase with some power of the window size 
\begin{equation}
\CN_c(l) \propto l^{d_f}. \label{massvsl}
\end{equation}
The exponent $d_f$ is the fractal dimension. 

\begin{figure}[tbph]
{ \hfill \fbox{%
\psfig{figure=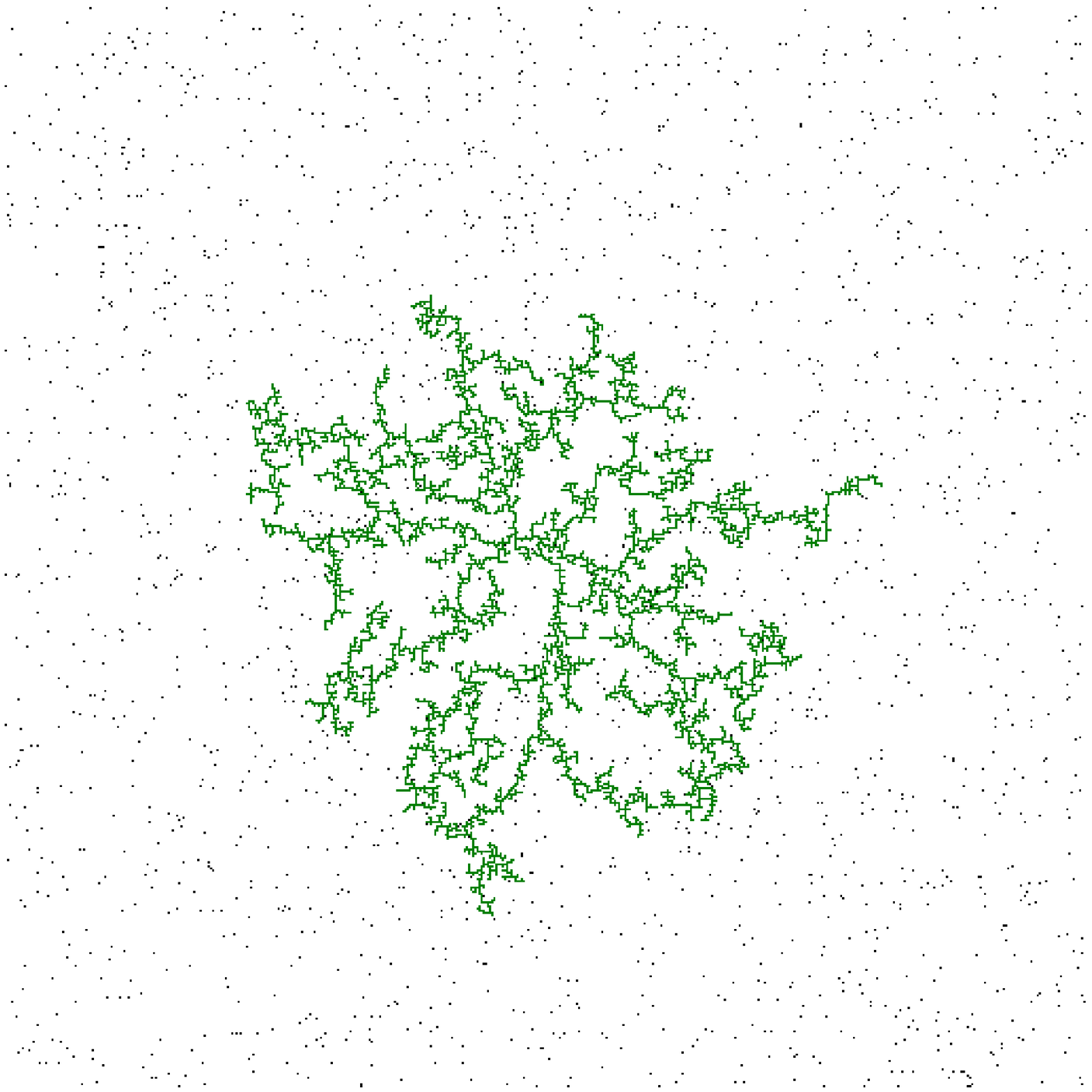,width=2.25in}}\hfill \fbox{%
\psfig{figure=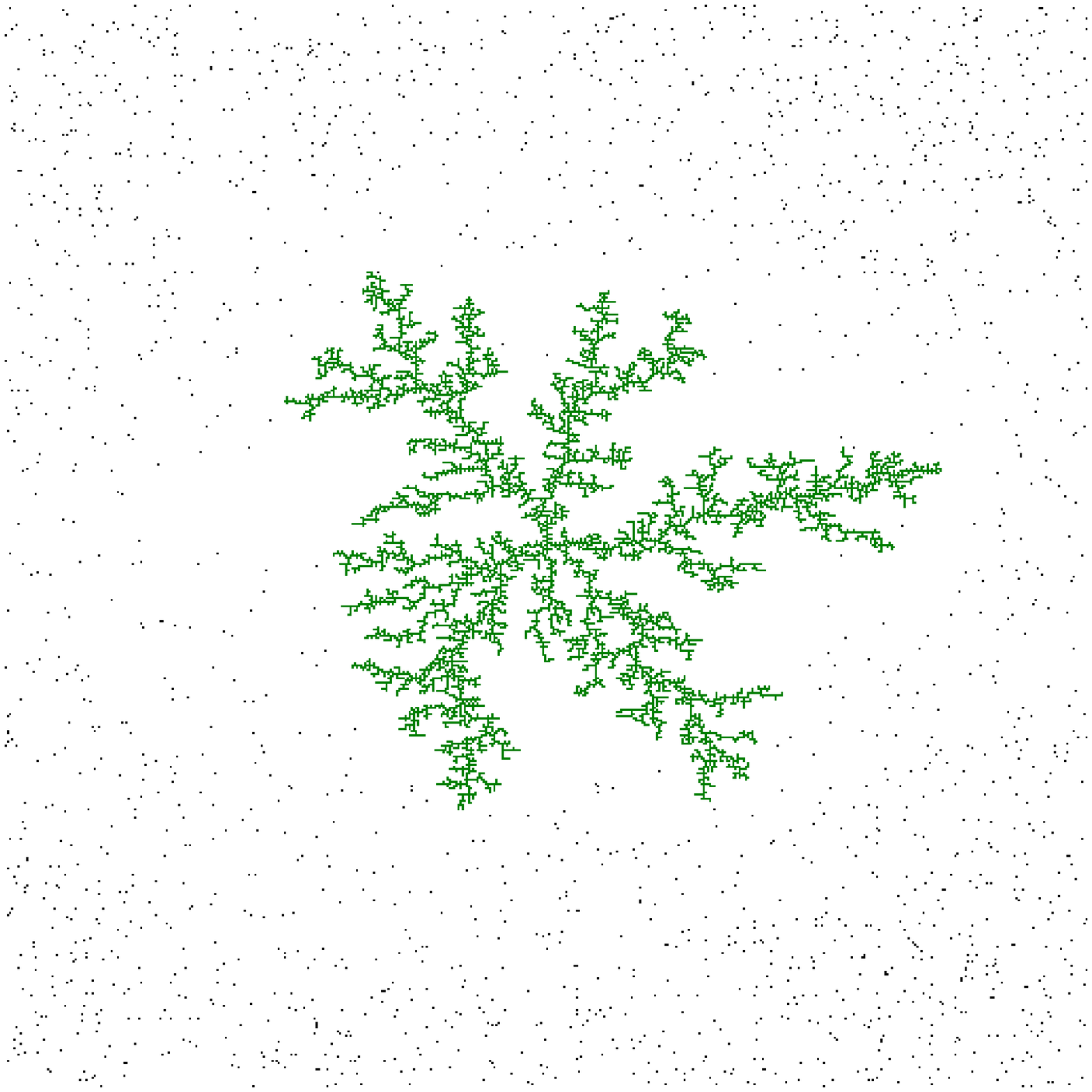,width=2.25in}}\hfill}%
\medskip\caption{Two growth clusters of the same mass, $\CN_c \sim
8270$. (a) A cluster grown via the RA model, pictured at time $t=
\tau_{\rm{T}}$, where $\tau_{\rm{T}}$ is the time for the heat bath
and gas-aggregate system to reach the same temperature. (b) A parallel DLA
cluster. Note the gas particles, which are shown as the small
dots. For the RA system the gas particles are distributed throughout
the space, yet for the DLA system very few gas particles penetrate the
region defined by the edges of the cluster.}  
\label{fig3.ravsdla}
\end{figure}

The RA cluster should initially resemble a parallel, irreversible DLA
cluster of equivalent mass.  Figure~\ref{fig3.ravsdla}(a) shows a typical RA
cluster for the $L=512$ system at the time $t=\tau_{\rm{T}}$, which is
the time for the mass of the RA cluster to stabilize at essentially
its final mass ($\CN_c \simeq 8270$). Figure~\ref{fig3.ravsdla}(b) shows a
typical DLA cluster of equivalent mass. Both systems were initialized
with a $4\%$ density of diffusing gas particles. The gas particles
still present at this stage of the evolution are shown as the
small dots in the figure. Note that for the RA system the gas particles
are distributed throughout the space, yet for the DLA system very few
gas particles penetrate the region defined by the edges of the
cluster.  The RA cluster has experienced enough annealing by the time
$t=\tau_{\rm{T}}$ to have a fractal dimension different than that of
the DLA cluster, yet the radii of both clusters are comparable and
approximately equal to a quarter of the length of the system
($r\approx L/4$). The RA cluster morphology is still far from its
final morphology.  

\begin{figure}[tbph]
\centerline{\psfig{figure=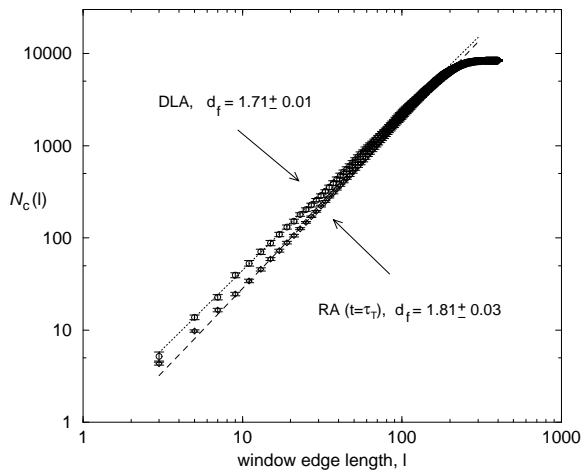,width=3in}}
\medskip\caption{The number of aggregate particles contained
in a box of length $l$, as a function of $l$.  The slope of the
line is the fractal dimension.  The top curve is for parallel
DLA clusters of mass $\CN_c \simeq 8270$. The
bottom curve is for the RA clusters sampled at $t=\tau_{\rm T}$.
Examples of these clusters are pictured in Fig.~\ref{fig3.ravsdla}.}  
\label{fig4.boxdf}
\end{figure}

Figure \ref{fig4.boxdf} shows the box-counting results obtained
for both models in the regime described above and pictured in
Fig.~\ref{fig3.ravsdla}.  The top curve is for DLA, the bottom for RA.  
Both models were implemented on a $L=512$ size system. The 
vertical axis is the mass contained in the window, $\CN_c(l)$, the
horizontal is the window size $l$. The curve for the DLA system is the
average of $10$ independent DLA clusters of mass $\CN_c \simeq
8270$.  The curve for the RA system is the average of ten
independent RA clusters sampled at time $t=\tau_{\rm{T}}$. The slope of the
curve corresponds to the fractal dimension and was determined via a 
linear least squares fit.  Consistent with past numerical studies of
DLA\cite{Stanley-dla-rev,HalsLeib-dla92}, 
we find that the fractal dimension for DLA clusters is $d_f^{\rm{DLA}}
= 1.71 \pm 0.01$ (for difficulties associated with determining the
fractal dimension of DLA see the detailed discussion in
Ref.~\cite{ErzanPietVesp-rmp95}). For the RA clusters the 
fractal dimension is $d_f^{\rm{RA}}(t=\tau_{\rm{T}}) = 1.81 \pm 0.03$.
A line with this slope is shown overlaying each respective
curve\cite{scaling-note}. 

\end{multicols}\setlength{\columnwidth}{\hsize}
\bigskip
\begin{multicols}{2}\setlength{\columnwidth}{\hsize}

\bigskip

The RA cluster is less dense than the DLA cluster in the area
immediately surrounding the initial aggregation site, however the
radii of both clusters are comparable. Several of the aggregate
particles in the RA cluster have annealed away from the center to
occupy the region between the center and the edge of the
cluster. Hence, at the time depicted in Fig.~\ref{fig3.ravsdla}, the
RA cluster has a higher fractal 
dimension than the DLA cluster.  The only constraints on the cluster are
the number of particles and the connectivity.  As there are more
ways to have connected clusters of a specified particle number in an
area of larger radius, the RA cluster evolves from the dense,
bushy DLA-like structure shown in Fig. \ref{fig3.ravsdla}, to a
tenuous structure which occupies more of the available lattice (with
the initial increase in fractal dimension being a transient
behavior).  We ultimately expect to observe a diffuse structure with
just a few meandering vines which can access more of the available
configuration space.  

\begin{figure}[tbph]
\centerline{\fbox{\psfig{figure=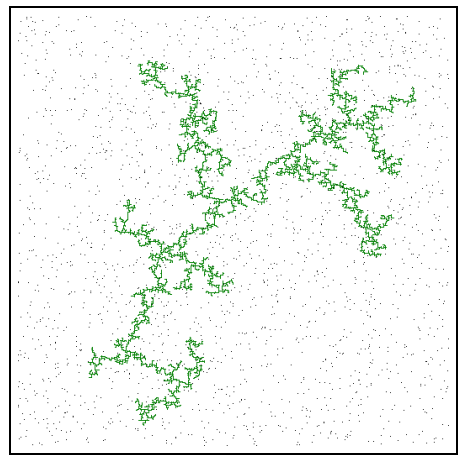,width=2.25in}}}
\medskip\caption{A growth cluster grown via the RA model,
pictured at time $t=80\tau_{\rm{T}}$. The fractal dimension
for this cluster ($d_f = 1.63 \pm 0.02$) has seemingly
reached the  asymptotic value.}
\label{fig5.ra-t80mil}
\end{figure}

As the time into the simulation advances, the density of the growth
aggregate decreases, the radius of the aggregate increases, and hence
the fractal dimension decreases.  Figure \ref{fig5.ra-t80mil} shows a
typical RA growth cluster at the time $t=80\tau_{\rm{T}}$ timesteps.
Note that the structure does resemble meandering vines.  Also the
radius of the cluster is comparable to half of the lattice size
($r\approx L/2$). 

\begin{figure}[tbph]
\centerline{\psfig{figure=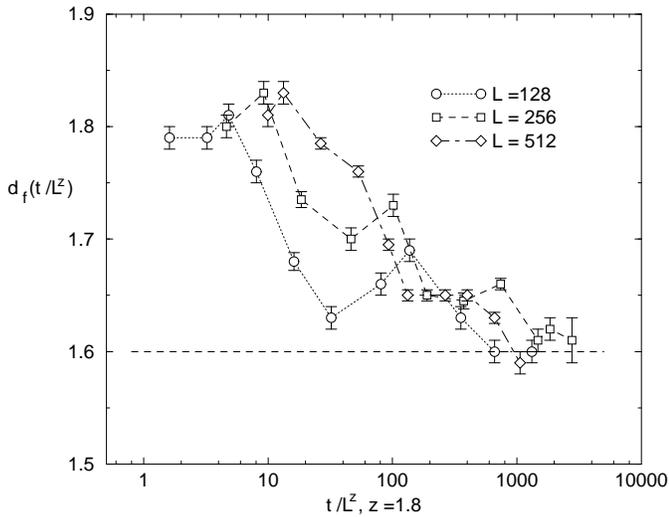,width=3.5in}}
\medskip\caption{The average fractal dimension
of the RA growth clusters as a function of
time into the simulation, for all three system sizes.}
\label{fig6.boxdfvstime}
\end{figure}

Figure \ref{fig6.boxdfvstime} is a plot of the average fractal dimension
for RA clusters as a function of time into the simulation, for all
three system sizes.  The measurements reported below are averages over
$5$ independent realizations for the $L=512$ system, $10$ independent
realizations for the $L=256$ system, and $10$ for the $L=128$ system
(i.e., averages over either $5$ or $10$ independently generated, large
clusters). The data points and standard errors shown in the plots 
are the average and standard error over the set of
independent realizations.  

The fractal dimension is initially very close to
the fractal dimension for DLA.  We then observe a slight increase in
the fractal dimension as the cluster center begins to anneal (an
example is the RA cluster shown in Fig.~\ref{fig3.ravsdla}), then a
gradual decrease in the fractal dimension until it converges upon an
equilibrium value. The  solid line is drawn to denote the equilibrium
value upon which results for the three system sizes are converging,
$d_f =1.6$.  Using Flory-type scaling arguments it has previously been
shown that a quenched branched polymer obeys the scaling relationship
$N \sim R^{d_Q}$, with $d_Q = [2(D+2)]/5$\cite{Daoud-etal-macromol83}.
Here $R$ represents the characteristic end-to-end distance of a
polymer, and $D$ the dimension of the space. $R$ can be taken in
direct analogy to $l$ in Eq. (\ref{massvsl}), which defines the
end-to-end distance of the window of interest. For $D=2$ the
exponent $d_Q = 1.6$. We should note that an exact result was obtained
for quenched polymers in $D=2$, $d_Q = 1.64$\cite{ParisiSour-PRL81},
which is slightly larger. Flory-type scaling has also been studied
for annealed branched polymers and the scaling exponent determined to
be $d_A = (3D+4)/7$\cite{GutGrosShak-macromol93}.  For $D=2$, $d_A =
1.43$.  One might expect to observe a crossover from quenched to
annealed behavior for the equilibrium RA growth clusters as we go
from the large to the small system sizes, but we did not see this for
the system sizes investigated.

Note that the time axis in Fig.~\ref{fig6.boxdfvstime} is rescaled by
$L^{1.8}$ in order to match that of Fig.~\ref{fig2.temp-eqm}.  Neither the
fractal dimension nor the equilibrium temperature exhibit finite size
effects as far as we can determine within the precision of our
measurements.  

\bigskip

\bigskip

\section{Discussion}

We have presented a microscopically reversible model which exhibits
macroscopic pattern formation.  In this model, entropy initially grows
rapidly with time, and then subsequently grows exceedingly
slowly---the slow relaxation can be characterized as a quasistatic
isothermal process.  The morphology of the aggregate formed by this
dynamics changes markedly with time, evolving from a pattern having a
conformation and fractal dimension similar to that of an irreversible
DLA system, to a pattern characteristic of a branched quenched random
polymer.

The RA model is an extension of the standard DLA model: 
we model the latent heat released when a gas particle aggregates in
addition to modeling the gas and crystal particles.  Since the
dynamics is local and microscopically reversible, we are realistically
modeling the flow of heat and the creation of entropy in this system
and thus we model the thermodynamic behavior of growing clusters. 

The model presented is simple and amenable to theoretical analysis.
Given the constraints that we set out, an even simpler model could be
constructed with just a single heat bath particle at each site, and a
single gas particle at each site.  In this case, diffusion would be
performed by block partitioning\cite{Margolus-feynlec}.  This simpler
model would have two fewer bits of persistent state at each lattice
site than the RA model, and would be slightly easier to analyze
theoretically.  It would, however, be less computationally efficient:
for a given lattice size, the volume available to the gas and heat
particles would be reduced, but the computational effort required by
each step of the simulation would be unchanged.

There are some simple variants of the RA model which might merit
study.  For example, we have only investigated situations in which the
temperature is set by the size of the final aggregate.  It would be
interesting to study morphology in situations where there is
independent control of temperature and aggregate size.  It would also
be interesting from a thermodynamic perspective to modify the model by
introducing a gas-crystal exclusion: gas particles would collide with
the aggregate, but not diffuse over it.  Thus there would be an
excluded volume for the gas particles, and the crystal would do work
on the gas as it grew.

\acknowledgements
The work at MIT was supported by DARPA contract number
DABT63-95-C-0130.  The authors would like to thank M. Smith,
A. Erzan, and Y. Bar-Yam for discussions useful in formulating this
model, as well as M. Kardar, B. Boghosian, and H. Mayer for helpful 
critiques.  

\end{multicols}\setlength{\columnwidth}{\hsize}
\bibliographystyle{unsrt}

\begin{thebibliography}{10}

\bibitem{Prig}
G.~Nicolis and I.~Progogine.
\newblock {\em Self-Organization in Nonequilibrium Systems}.
\newblock Wiley-Interscience, New York, 1977.

\bibitem{cambook}
T.~Toffoli and N.~H. Margolus.
\newblock {\em Cellular Automata Machines: A New Environment for Modeling}.
\newblock MIT Press, Cambridge, MA, 1987.

\bibitem{Margolus-feynlec}
N.~H. Margolus.
\newblock Crystalline computation, to appear.
\newblock In A.~Hey, editor, {\em Feynman and Computation}. Addison-Wesley,
  1998.

\bibitem{dsouza-same3d}
R.~M. D'Souza.
\newblock Reversible pattern formation in {Ising}-like cellular automata.
\newblock {\em manuscript in preparation}.

\bibitem{Vold}
M.~J. Vold.
\newblock A numerical approach to the problem of sediment volume.
\newblock {\em Jour. Colloid Sci.}, 14:168--174, 1959.

\bibitem{KKRSOS}
J.~M. Kim and J.~M. Kosterlitz.
\newblock Growth in a restricted solid-on-solid model.
\newblock {\em Phys. Rev. Lett.}, 62(19):2289--2292, 1989.

\bibitem{Pearson_turing}
J.~E. Pearson.
\newblock Complex patterns in a simple system.
\newblock {\em Science}, 261:189--194, 1993.

\bibitem{Ben-Jac_bacsnowflakes}
E.~Ben-Jacob {\em et al.}
\newblock Response of bacterial colonies to imposed anisotropy.
\newblock {\em Phys. Rev. E}, 53(2):1835--1843, 1996.

\bibitem{Eden-model}
M.~Eden.
\newblock A two-dimensional growth process.
\newblock In J.~Neyman, editor, {\em Proceedings of the Fourth Berkeley
  Symposium on Mathematical Statistics and Probability, Volume IV}, pages
  223--239, 1961.

\bibitem{WittSand-dla}
T.~A. Witten and L.~M. Sander.
\newblock {Diffusion-Limited Aggregation}, a kinetic critical phenomenon.
\newblock {\em Phys. Rev. Lett.}, 47(19):1400--1403, 1981.

\bibitem{Landauer-diss61}
R.~Landauer.
\newblock Irreversibility and heat generation in the computing process.
\newblock {\em IBM Jour. Res. Dev.}, 3:183--191, 1961.

\bibitem{Benn-thermo}
C.~H. Bennett.
\newblock Thermodynamics of computation.
\newblock {\em Int. J. of Theor. Phys.}, 21:905--940, 1982.

\bibitem{mpf-foresight}
M.~P. Frank and {T. F. Knight, Jr.}
\newblock Ultimate theoretical models of nanocomputers.
\newblock {\em Nanotechnology}, 9:162--176, 1998.

\bibitem{Creutz-annphys86}
M.~Creutz.
\newblock Deterministic ising dynamics.
\newblock {\em Annals of Physics}, 167:62--72, 1986.

\bibitem{Pomeau-jphysa84}
Y.~Pomeau.
\newblock Invariant in cellular automata.
\newblock {\em J. Phys. A}, 17:L415--L418, 1984.

\bibitem{Vichniac-physd84}
G.~Vichniac.
\newblock Simulating physics with cellular automata.
\newblock {\em Physica D}, 10(1/2):96--116, 1984.

\bibitem{HPP-pra76}
J.~Hardy, O.~de~Pazzis, and Y.~Pomeau.
\newblock Molecular dynamics of a classical lattice gas: transport properties
  and time correlation functions.
\newblock {\em Phys. Rev. A}, 13(5):1949--1961, 1976.

\bibitem{FHP-prl86}
U.~Frisch, B.~Hasslacher, and Y.~Pomeau.
\newblock Lattice-gas automata for the {Navier-Stokes} equation.
\newblock {\em Phys. Rev. Let.}, 56(14):1505--1508, 1986.

\bibitem{Lebowitz}
J.~L. Lebowitz.
\newblock Microscopic reversibility and macroscopic behavior: Physical
  explanations and mathematical derivations.
\newblock In J.~J. Brey, J.~Marro, J.~M. Rubi, and M.~San Miguel, editors, {\em
  Lecture Notes in Physics}. Springer, 1995.

\bibitem{LevesVerl-jstatphys93}
D.~Levesque and L.~Verlet.
\newblock Molecular-dynamics and time reversibility.
\newblock {\em J. Stat. Phys.}, 72(3-4):519--537, 1993.

\bibitem{nhm-physlike}
N.~H. Margolus.
\newblock Physics-like models of computation.
\newblock {\em Physica D}, 10:81--95, 1984.

\bibitem{Stanley-dla-rev}
H.~E. Stanley.
\newblock Fractals and multifractals: The interplay of physics and geometry.
\newblock In A.~Bunde and S.~Havlin, editors, {\em Fractals and Disordered
  Systems}, New York, 1996. Springer-Verlag.

\bibitem{Langer-rmp80}
J.~S. Langer.
\newblock Instabilities and pattern formation in crystal growth.
\newblock {\em Rev. Mod. Phys.}, 52(1):1--28, 1980.

\bibitem{ErzanPietVesp-rmp95}
A.~Erzan, L.~Pietronero, and A.~Vespignani.
\newblock The fixed-scale transformation approach to fractal growth.
\newblock {\em Rev. Mod. Phys.}, 67(3):545--604, 1995.

\bibitem{Voss-jstatp84}
R.~F. Voss.
\newblock Multiparticle fractal aggregation.
\newblock {\em J. Stat. Phys.}, 36(5/6):861--872, 1984.

\bibitem{Naga-jphysjpn92}
T.~Nagatani.
\newblock Unsteady diffusion-limited aggregation.
\newblock {\em J. Phys. Soc. Jpn.}, 61(5):1437--1440, 1992.

\bibitem{Smith-phd}
M.~A. Smith.
\newblock {\em Cellular Automata Methods in Mathematical Physics}.
\newblock PhD thesis, Massachusetts Institute of Technology, May 1994.

\bibitem{ChopDroz-diff}
B.~Chopard and M.~Droz.
\newblock Cellular automata model for the diffusion equation.
\newblock {\em J. Stat. Phys.}, 64(3/4):859--892, 1991.

\bibitem{cam8-waterloo}
N.~H. Margolus.
\newblock {\sc CAM-8}: a computer architecture based on cellular automata.
\newblock In A.~Lawniczak and R.~Kapral, editors, {\em Pattern Formation and
  Lattice-Gas Automata}. American Mathematical Society, 1996.

\bibitem{Bmb-renorm}
B.~M. Boghosian and W.~Taylor.
\newblock Correlations and renormalization in lattice gases.
\newblock {\em Phys. Rev. E}, 52(1):510--554, 1995.

\bibitem{Hassl-discretefluids}
B.~Hasslacher.
\newblock Discrete fluids.
\newblock {\em Los Alamos Science}, 1987.

\bibitem{Huang-statmech}
K.~Huang.
\newblock {\em Statistical Mechanics, 2nd ed.}
\newblock Wiley, New York, 1987.

\bibitem{HalsLeib-dla92}
T.~C. Halsey and M.~Leibig.
\newblock Theory of branched growth.
\newblock {\em Phys. Rev. A}, 46(12):7793--7809, 1992.

\bibitem{scaling-note}
Note that in the limit where the cluster size approaches infinity, a cluster
  grown via a parallel implementation of {DLA} will exhibit a crossover from
  fractal to two-dimensional. in our simulations of parallel {DLA} the initial
  density of gas particles is dilute enough, and the cluster size small enough,
  that we do not see any crossover effects and we obtain smooth scaling curves
  as shown in {Fig}. 4.

\bibitem{Daoud-etal-macromol83}
M.~Daoud, P.~Pincus, W.~H. Stockmayer, and T.~Witten.
\newblock Phase separation in branched polymer solutions.
\newblock {\em Macromolecules}, 16(12):1833--1839, 1983.

\bibitem{ParisiSour-PRL81}
G.~Parisi and N.~Sourlas.
\newblock Critical behavior of branched polymers and the {Lee-Yang}
  singularity.
\newblock {\em Phys. Rev. Lett.}, 46(14):871--874, 1981.

\bibitem{GutGrosShak-macromol93}
A.~M. Gutin, A.~Yu. Grosberg, and E.~I. Shakhnovich.
\newblock Polymers with annealed and quenched branchings belong to different
  universality classes.
\newblock {\em Macromolecules}, 26(6):1293--1295, 1993.

\end{thebibliography}

\end{document}